
\magnification= \magstep1
\baselineskip=14pt
\nopagenumbers
\newcount\knum
\knum=1
\newcount\uknum
\uknum=1
\newcount\znum
\znum=1
\newcount\notenumber

\def\Za{\the\uknum.\the\znum \global\advance\znum by 1}
\def\ukneu{\znum=1\global\advance\uknum by 1}
\font\tbfontt=cmbx10 scaled \magstep0
\font\gross=cmbx10 scaled \magstep2

\font\mittel=cmbx10 scaled \magstep1
\font\ittel=cmr10 scaled \magstep1

\font\TT=cmcsc10 scaled \magstep0

\font\eightrm=cmr8

\font\eightit=cmti8 scaled \magstep0
\def\sqr#1#2{{\vcenter{\vbox{\hrule height.#2pt\hbox{\vrule width.#2pt
height#1pt \kern#1pt \vrule width.#2pt}\hrule height.#2pt}}}}
\def\square{\mathchoice\sqr34\sqr34\sqr{2.1}9\sqr{1.5}9}
\def\QED{\smallskip\rightline{$\square$ \quad\qquad\null}}

\def\Theorem{\vskip 0.3 true cm\sl {\TT Theorem \Za.}{ }}

\def\Lemma{\vskip 0.3 true cm\sl {\TT Lemma \Za.}{ }}
\def\Proof{\TT Proof:{ }\rm}

\def\kz{{\rm C\hskip-4.0pt\vrule height4.3pt\hskip5.3pt}}  

\def\eins{{\rm 1\hskip -2pt I}}

\def\lim{\mathop{\rm lim}}

\def\End{{\hbox{\rm End}}}

\def\dim{{\hbox{\rm dim}}}

\parindent=0pt
\font\eightrm                = cmr8
\font\eightsl                = cmsl8
\font\eightsy                = cmsy8
\font\eightit                = cmti8

\font\eighti                 = cmmi8
\font\eightbf                = cmbx8
\def\petit{\def\rm{\fam0\eightrm}
\textfont0=\eightrm 
 \textfont1=\eighti 
 \textfont2=\eightsy 
 \def\it{\fam\itfam\eightit}
 \textfont\itfam=\eightit
 \def\sl{\fam\slfam\eightsl}
 \textfont\slfam=\eightsl
 \def\bf{\fam\bffam\eightbf}
 \textfont\bffam=\eightbf 
 \normalbaselineskip=9pt
 \setbox\strutbox=\hbox{\vrule height7pt depth2pt width0pt}
 \normalbaselines\rm}
\newdimen\refindent
\def\begref{\vskip1cm\bgroup\petit
\setbox0=\hbox{[Bi,Sc,So]o }\refindent=\wd0
\let\sl=\rm\let\INS=N}

\def\ref#1{\filbreak\if N\INS\let\INS=Y\vbox{\noindent\tbfontt
Literatur\vskip1cm}\fi\hangindent\refindent
\hangafter=1\noindent\hbox to\refindent{#1\hfil}\ignorespaces}

\long\def\fussnote#1#2{{\baselineskip=9pt
\setbox\strutbox=\hbox{\vrule height 7pt depth 2pt width 0pt}%
\petit\noindent\footnote{\noindent #1}{#2}}}
\rightline{hep-th/9503152}
\rightline{CPT-94/P.3106}
\rightline{Mannheimer Manuskripte 181}
\rightline{November 1994}
\vskip 4cm
\phantom{prelim.version}
\centerline{\gross A generalized Lichnerowicz formula,}
\smallskip
\centerline{\gross the Wodzicki Residue
and Gravity }
\vskip 1cm

\centerline{\ittel Thomas Ackermann\footnote{$^1$}{\eightrm
e-mail: ackerm@euler.math.uni-mannheim.de} and
J\"urgen Tolksdorf\footnote{$^2$}{\eightrm
e-mail: tolkdorf@cptsu4.univ-mrs.fr}\ \footnote{${^\star}$}{
\eightrm Supported
by the European Communities,
contract no.\hskip -0.3cm \hbox{\petit\noindent\rm ERB 401GT 930224}}}
\vskip 0.5cm
\centerline{\vbox{\hsize=4.7 true in \rm\noindent
$^1$\ Fakult\"at f\"ur Mathematik \& Informatik, Universit\"at
Mannheim \break \phantom{$^1$} D-68159 Mannheim, F.R.G.}}
\smallskip
\centerline{\vbox{\hsize=4.7 true in \rm\noindent
$^2$\ Centre de Physique Th${\acute {\rm e}}$orique, CNRS Luminy,
Case 907 \phantom{aaaaaa} \break
\phantom{$^2$} F-13288 Marseille Cedex 9, France }}
\vskip 2.0cm
\centerline{\vbox{\hsize=5.0 true in \petit\noindent
\bf Abstract.
\rm We prove a generalized version of the well-known
Lichnerowicz formula for the square of the most general Dirac
operator $\widetilde{D}$\ on an even-dimensional spin manifold
associated to a metric connection $\widetilde{\nabla}$. We use
this formula to compute the subleading term $\Phi_1(x,x,
\widetilde{D}^2)$\ of the heat-kernel expansion of $\widetilde{D}^2$.
The trace of this term plays a key-r$\hat {\petit\rm o}$le in
the definition of a (euclidian) gravity action in the context of
non-commutative geometry. We show that this gravity action can be
interpreted as defining a modified euclidian Einstein-Cartan
theory.}}
\vskip 0.8cm
Keywords: \it non-commutative geometry, Lichnerowicz formula,
Dirac operator, heat-kernel-\phantom{Keywords: }expansion,
Wodzicki residue,
gravity\rm\vfil\break
\advance\hsize by -0.5true in
\advance\vsize by -0.9true in
\advance\hoffset by 0.45in
\advance\voffset by 0.7in
\pageno=1
\topskip=1.2cm plus 0.2cm
\parindent=0pt
{\mittel 1. Introduction}
\vskip 0.7cm
When attempting to quantize the electron
in 1928, Dirac
introduced a first-order operator the square of which is the
so-called wave-operator (d'Alembertian operator).
Lateron, in the hands of mathematicians
generalizations of this operator, called `Dirac operators' -
evolved into an important tool of modern mathematics, occuring
for example in index theory, gauge theory, geometric
quantisation etc.
\smallskip
More recently, Dirac operators have assumed a significant
place in Connes' non-commutative geometry [C] as the
main ingredient in the definition of a K-cycle. Here they
encode
the geometric structure of the underlying non-commutative
`quantum-spaces'. Thus disguised,
Dirac operators
re-enter modern physics, since
non-com\-mutative geometry can be used, e.g. to derive the action of
the standard model of
elementary particles, as shown in [CL] and [K1]. Initially,
it remains unclear wether it was possible to also
derive the Einstein-Hilbert action of gravity using this approach.
And again, it was a
Dirac operator which proved to be
the key to answer
this question. According to Connes [C],
the `usual' Dirac operator
$D$\ on the spinor bundle $S$\ associated to the Levi-Civita connection
on a four-dimensional spin manifold $M$\ is linked to
the euclidian Einstein-Hilbert gravity action via the Wodzicki
residue of the inverse of $D^2$. This was shown in detail in [K2].
A further question that naturally arises is the
the dependence of this result
from the chosen Dirac operator $D$. In other words, does
$Res(\widetilde{D}^{-2})$\ change if $\widetilde{D}$\ is
a Dirac operator on $S$\ different from the `usual' one ?
\medskip
In this paper we answer this question affirmatively. Moreover, in
section 3,
we compute the lagrangian of an appropriatly
defined gravity action
$${I_{GR}(\widetilde{D}):=-{2\over 2^n (2n-1)}\;
Res(\widetilde{D}^{-2n+2})}\eqno(1.1)$$
for the
most general Dirac operator $\widetilde{D}$\
associated to a metric connection
$\widetilde{\nabla}$\
on a compact spin manifold $M$\ with $\dim\;M=2n\ge 4$.
We proceed as follows:
According to
the main theorem of [KW], there is a relation between
the Wodzicki residue
$Res({\hat\triangle}^{-n+1})$\
of a generalized laplacian
$\hat\triangle$\ on a hermitian bundle $E$\ over $M$\ and
$\Phi_1(x,x,{\hat\triangle})$, which denotes the
the subleading term of
its the heat-kernel expansion.
It is well-know that, given any generalized Laplacian
$\hat\triangle$\ on $E$, there exists a
connection ${{\hat\nabla}^E}$\ on $E$\ and a section $F$\ of
the endomorphism bundle $\End(E)$, such that $\hat\triangle$\ decomposes
as
$${{\hat\triangle}=\triangle^{{\hat\nabla}^E} + F.}\eqno(1.2)$$
However, this has a slight flaw. The decomposition (1.2)
neither provides any
method
to construct the connection $\hat\nabla^E$\ nor the endomorphism
$F$\ explicitly. Nevertheless, it is exactly this endomorphism
$F\in \Gamma(\End E)$\ which fully determines the subleading term
$\Phi_1(x,x,{\hat\triangle})$\ of the heat-kernel-expansion
of ${\hat\triangle}$\ (cf. [BGV]). Thus the
the problem of computing $Res({\hat\triangle}^{-n+1})$\
is transformed into the problem of
computing $F$.
\smallskip
For an arbitrary generalized laplacian, this
might prove to be difficult. However, in the case where $E=S$\ and
the generalized laplacian $\hat \triangle$\ is the square
${\widetilde{D}^2}$\ of a Dirac operator associated
to an arbitrary metric connection $\widetilde{\nabla}$\ on $TM$,
a constructive
version of (1.2) can be proved. This will be shown in
section 2 . Because
of its
close relationship to the well-known Lichnerowicz formula (cf. [L])
we call our decomposition formula a `generalized Lichnerowicz
formula'. We understand it as being intrinsic to the
the Dirac operators studied in this paper.
\smallskip
As already mentioned, we will
compute the
lagrangian of (1.1) in section 3, using our generalized Lichnerowicz
formula
as the main technical tool. From a physical point of view, this
lagrangian can be interpreted as defining a modified (euclidian)
Einstein-Cartan theory.
\vskip 1.2cm
{\mittel 2. A generalized Lichnerowicz formula}
\vskip 0.7cm\ukneu
Let $M$\ be a spin manifold with $\dim\; M=2n$\
and let us denote its Riemannian metric by $g$..
The Levi-Civita connection ${\nabla\colon \Gamma(TM)\rightarrow
\Gamma(T^*M\otimes TM)}$\ on $M$\ induces a
a connection ${\nabla^S\colon
\Gamma(S)\rightarrow \Gamma(T^*M\otimes S)}$\ on the spinor bundle
$S$\ which is compatible with the hermitian metric
$<\;\cdot\;,\;\cdot\;>_S$\ on $S$. By adding an
additional torsion term $t\in \Omega^1(M,\End TM)$\
we obtain a new covariant derivative
$${\widetilde{\nabla}:=\nabla + t}\eqno(2.1)$$
on the tangent bundle $TM$. Since $t$\ is really
a one-form on $M$\ with values in the bundle of skew endomorphism
$Sk(TM)$\ (cf. [GHV]), $\widetilde{\nabla}$\ is in fact
compatible with the Riemannian metric $g$\ and
therefore also induces a connection
$\widetilde{\nabla}^S=\nabla^S+T$\ on the
spinor bundle. Here $T\in \Omega^1(M,\End S)$\ denotes the
`lifted' torsion term $t\in \Omega^1(M,\End TM)$.
However, in general this induced connection $\widetilde{\nabla}^S$\
is neither compatible with the hermitian metric $<\;\cdot\;,\;\cdot\;
>_S$\ nor compatible with the Clifford action on $S$.
With respect to
a local orthonormal frame
${\{e_a\}_{1\le a\le 2n}}$\ of $TM\vert_{U\subset M}$\ we have
$$\matrix{&\nabla_c\; e_b ={\omega^a}_{bc} e_a
&t:={t^a}_{bc}\; e_a\otimes e^b\otimes e^c \cr
\noalign{\smallskip}
&\nabla^S s_l ={{\petit 1}\over {\petit 4}}\; \gamma^a\gamma^b s_l
\otimes \omega_{abc} e^c
&\widetilde{\nabla}^S s_l ={{\petit 1}\over {\petit 4}}\;
\gamma^a\gamma^b s_l\otimes
\bigl(
\omega_{abc} + t_{abc}\bigr)e^c \cr}$$
where ${{\omega^a}_{bc}}$\ denotes the components of the Levi-Civita
connection, ${\{e^a\}_{1\le a\le 2n}}$\ the corresponding
dual frame
of ${\{e_a\}_{1\le a\le 2n}}$\
and ${\{s_l\}_{1\le l\le {\petit \rm dim }\;
S}}$\ a local frame of $S\vert_U$. Nnote that
we use the
following conventions
$${\{\gamma^a,\gamma^b\}=-2\eta^{ab},\hskip 2cm [\gamma^a,\gamma^b]=
2\gamma^{ab}}$$
for the representation $\gamma\colon C_\kz(T^*M)\rightarrow \End\;S $\
of the complexified Clifford algebra of $T^*M$\ on the spinor bundle.
\smallskip
We now define
by
${\widetilde{D}:=\gamma^\mu\widetilde{\nabla}_\mu^S}$\
a first order operator
${\widetilde{D}\colon \Gamma(S)\rightarrow
\Gamma(S)}$\ associated to the metric connection
${\widetilde{\nabla}}$. Since $\widetilde{D}$\ satisfies the
relationship
${[\widetilde{D}, f]=\gamma^\mu\;{\partial f\over \partial
x^\mu}}$\ for all $f\in C^\infty(M)$\
this operator $\widetilde{D}$\ is a Dirac operator, i.e.
its square $\widetilde{D}^2$\ is a generalized laplacian
(cf. [BGV]). Note that $\widetilde{D}$\ is also the most general
Dirac operator on the spinor bundle $S$\ corresponding to
a metric connection $\widetilde{\nabla}$\ on $TM$.
\smallskip
According to the well-known Ricci lemma (cf. [GHV]) there is a one-to-one
correspondence bet\-ween metric connections on $TM$\ and
the elements of $\Omega^1(M,Sk(TM))$.
Consequently, the set of
all such Dirac operators
acting on sections of the spinor bundle $S$\ over $M$\ is
parametrized by $t\in \Omega^1(M,Sk(TM))$.
\smallskip
For the square of the Dirac operator $\widetilde{D}$\
we get the following
standard decomposition
$${\widetilde{D}^2= -g^{\mu\nu}\widetilde{\nabla}^S_\mu\widetilde{
\nabla}^S_\nu
+ \gamma^\mu[\widetilde{\nabla}^S_\mu,\gamma^\nu]\widetilde{\nabla}^S_\nu
+ {1\over 2}\;\gamma^\mu\gamma^\nu [\widetilde{\nabla}^S_\mu,
\widetilde{\nabla}^S_\nu].}\eqno(2.2)$$
If $t=0$, which means that $\widetilde{\nabla}$\ is identical with
the Levi-Civita connection $\nabla$, equation (2.2)
is the first step to compute
the well-known Lichnerowicz formula of $D^2$, cf. [L].
Note that none of the first two terms of (2.2) is covariant in itself
but only their sum. Using (2.1) we can, however, rearrange
the decomposition (2.2)
such that each term is manifest covariant. Moreover the
derivations in the decomposition will then be
arranged according to their degree:
\Lemma Let $M$\ be a spin mannifold, $\nabla$\ the Levi-Civita
connection on $TM$\ and $\widetilde{\nabla}$\ defined by
${\widetilde{\nabla}:=\nabla + t}$\ with $t\in \Omega^1(M, Sk(TM))$.
Then the square $\widetilde{D}^2$\
of the Dirac operator
$\widetilde{D}$\ on the spinor bundle $S$\
associated to $\widetilde{\nabla}$\
decomposes as
$${\widetilde{D}^2=\triangle^\nabla - B^\mu \nabla^S_\mu +F^\prime}
\eqno(2.3)$$
where $B\colon \Gamma(TM\otimes\End S)$\ and $F^\prime\in \Gamma(\End S)$\
are defined by
$$\eqalignno{ B^a &:= 2T^a-\gamma^c[T_c,\gamma^a]
&(2.4)\cr
F^\prime &:={{\petit\rm 1}\over {\petit\rm 4}}\;R^\nabla \cdot
\eins_{{\petit
\rm End}\;S}
+
\gamma^a\gamma^b\bigl(^\prime\nabla_a T_b\bigr)
+\gamma^a T_a\gamma^b T_b. &(2.5) \cr}$$
Furthermore $R^\nabla$\ denotes the scalar curvature
and ${\triangle^\nabla:=\eta^{ab}(\nabla^S_a\nabla^S_b -\nabla^S_{
\nabla_a e_b})}$\ the
horizontal laplacian on the spinor bundle corresponding to
the Levi-Civita connection $\nabla$\ with respect
to a local orthonormal frame
${\{e_a\}_{1\le a\le 2n}}$.
\smallskip
\Proof By inserting ${\widetilde{\nabla}^S_\mu=\nabla^S_\mu + T_\mu}$\
in (2.2) and using the compatibility of the connection $\nabla^S$\ with the
Clifford action, so that ${[\nabla^S_\mu,\gamma^\sigma]=-\gamma^\nu
\Gamma^\sigma_{\nu\mu}}$\
we get
$$\eqalign{\widetilde{D}^2 &=\triangle^\nabla +{{\petit\rm 1}\over
{\petit\rm 2}}\;
\gamma^{\mu\nu}\;[\nabla^S_\mu,\nabla^S_\nu]-
g^{\mu\nu}\bigl([\nabla^S_\mu,T_\nu]
-\Gamma^\sigma_{\nu\mu}T_\sigma\bigr) \cr
&\ \ -g^{\mu\nu}T_\mu T_\nu
-g^{\mu\nu}\bigl(2 T_\mu\nabla^S_\nu\bigr) +
\gamma^\mu[T_\mu,\gamma^\nu]
\nabla^S_\nu +{{\petit\rm 1}\over {\petit 2}}\;\gamma^{\mu\nu}\;
[T_\mu,T_\nu] \cr
&\ \ +\gamma^{\mu\nu}[\nabla^S_\mu,T_\nu] +\gamma^\mu[T_\mu,\gamma^\nu]
T_\nu.\cr}\eqno(2.6)$$
With the help of the Clifford relation $\gamma^\mu\gamma^\nu +
\gamma^\nu\gamma^\mu=-2g^{\mu\nu}$\ and the
first Bianchi identity ${R_{jkli}+R_{klij}+R_{lijk}=0}$\ one can
identify the second term in (2.6) with the `usual' Lichnerowicz term:
$${{{\petit\rm 1}\over {\petit\rm 2}}\;
\gamma^{\mu\nu}\;[\nabla^S_\mu,\nabla^S_\nu]=
{{\petit\rm 1}\over {\petit\rm 4}}\;R^\nabla\cdot \eins_S.}$$
If we write ${\gamma^\mu[T_\mu,\gamma^\nu]\nabla_\nu=
g^{\mu\nu}\gamma^\sigma[T_\sigma,\gamma_\mu]\nabla_\nu}$,
we see that $B^\nu$\ is given by the sum of
the fifth together with the sixth term on the
right-hand-side. Furthermore we have the identities
$$\eqalign{-g^{\mu\nu}\bigl([\nabla^S_\mu,T_\nu]-\Gamma^\sigma_{
\mu\nu}T_\sigma\bigr)
= -g^{\mu\nu}\bigl((\nabla^{{\petit\rm End}\;S}_\mu T_\nu)-
\Gamma^\sigma_{\mu\nu}T_\sigma\bigr) &=
-g^{\mu\nu}(^\prime\nabla_\mu T_\nu)\cr
\gamma^{\mu\nu}[\nabla^S_\mu,T_\nu] =\gamma^{\mu\nu}\bigl(
(\nabla^{{\petit\rm End}\;S}_\mu T_\nu) -\Gamma^\sigma_{\mu\nu}T_\sigma
\bigr) &=
\gamma^{\mu\nu}(^\prime\nabla_\mu T_\nu).\cr}\eqno(2.7)$$
Here ${^\prime\nabla\colon \Gamma(\End S\otimes T^*M)\rightarrow
\Gamma^(T^*M\otimes\End S\otimes T^*M)}$\ denotes the
induced connection ${^\prime\nabla:=\nabla^{{\petit\rm End}\; S}\otimes
\eins_{T^*M}+ \eins_{{\petit\rm End}\; S}\otimes \nabla}$\ on
the tensor bundle $\End\;S\otimes T^*M$. Because $\nabla$\ respects
the Clifford relation this means that
$${^\prime\nabla_\mu T_\nu={{\petit\rm 1}\over {\petit\rm 4}}\;
\gamma^{ab}\nabla^{T^*M\otimes T^*M\otimes T^*M}_\mu t_{ab\nu}
\equiv
{{\petit\rm 1}\over {\petit\rm 4}}
\;\gamma^{ab} t_{ab\nu;\mu.}}$$
Due to
the fact that
${\gamma^{\mu\nu}-g^{\mu\nu}=\gamma^\mu\gamma^\nu}$\ we obtain our
result.
\QED
\smallskip
It is well-know (see [BGV]) that given any generalized Laplacian
$\hat\triangle$\ on a hermitian bundle $E$\ over $M$, there exists a
connection ${{\hat\nabla}^E}$\ on $E$\ and a section $F$\ of
the endomorphism bundle $\End(E)$, such that $\hat\triangle$\ decomposes as
$${{\hat\triangle}=\triangle^{{\hat\nabla}^E} + F.}\eqno(2.8)$$
As we have mentioned before, this statement does not offer
any possibility of calculating the endomorphism $F$\
explicitly. Since it can be shown (cf. [BGV]), however, that
$${\Phi_1(x,x,\hat\triangle)={1\over 6}\;R^\nabla\cdot \eins_E - F,}$$
it is evident that
$F$\ plays a leading r${\hat{\rm o}}$le in the
computation of the subleading term
$\Phi_1(x,x,{\hat\triangle})$\
in the asymptotic expansion of the heat-kernel of
$\hat\triangle$. Moreover, by the main theorem
of [KW]\fussnote{${^{(1)}}$}{We denote by $*$\ the
Hodge-star operator associated to the Riemannian metric $g$.}
$${Res\bigl({\hat\triangle}^{-n+1}\bigr)= {2n-1 \over 2}\;
\int_M\; *\;tr\bigl(\Phi_1(x,x,{\hat\triangle})
\bigr),}\eqno(2.9)$$
this endomorphism $F$\ also determines the Wodzicki residue
of ${\hat\triangle}^{-n+1}$\ which defines
gravity actions in the case of ${\hat \triangle}=\widetilde{D}^2$.
\smallskip
We shall now prove a theorem which enables us to compute
$F$\ explicitly in the case of
$E=S$\ and ${{\hat\triangle}:=\widetilde{
D}^2}$.
\Theorem Let the hypotheses be the same as in lemma 2.1 and let
${{\hat\triangle}=\widetilde{D}^2}$\ be
the square of the Dirac operator
${\widetilde{D}}$\ associated to $\widetilde{\nabla}$. Then
the covariant derivative ${{\hat\nabla}^S}$\
and the endomorphism $F\in \Gamma(\End S)$\ in the decomposition (2.8)
are defined as follows:
$$\eqalignno{ {\hat\nabla}^S &:=\nabla^S + {\hat T} &(2.10)\cr
                F &:= F^\prime + \aleph. &(2.11) \cr}$$
With respect to a local orthonormal frame
${\{e_a\}_{1\le a\le n}}$\ of $TM$,
${{\hat T}\in \Omega^1(M,\End S)}$\ and $\aleph\in
\Gamma(\End S)$\ are explicitly given by
$$\eqalignno{{\hat T}_a &=T_a-{1\over 2}\;\gamma^b [T_b,\gamma_a]
&(2.12) \cr
\aleph &= ^\prime\!\nabla_a{\hat T}^a + {\hat T}_a {\hat T}^a,
&(2.13)\cr}$$
where ${F^\prime\in \Gamma(\End S)}$\ is the endomorphism (2.5) of
lemma 2.1 .
\smallskip
\Proof The main ingredient of this proof
is the global decomposition formula
(2.3) of
$\widetilde{D}^2$\ as given in lemma 2.1 . Concerning the case of
${\hat\triangle}=\widetilde{D}^2$, equation (2.3) is but an
alternative version of (2.8). We can therefore prove the theorem
by inserting
(2.10), (2.11), (2.12) and (2.13)
into equation (2.8) which then is identical with (2.3).
\QED
\smallskip
Thus we obtain the following formula for the square of the
Dirac Operator ${\widetilde{D}}$:
$$\eqalign{\widetilde{D}^2=&\triangle^{{\hat\nabla}^S} +
{{\petit\rm 1}\over {\petit\rm 4}}\; R\cdot \eins_S +
\gamma^{ab}\bigl(^\prime\nabla_a T_b\bigr)
+ {1\over 2}\;\gamma^{ab}[T_a,T_b]\cr
&\ \ \  +{1\over 2}\; [\gamma^a[T_a,\gamma^b],T_b] -
{1\over 2}\;\gamma^b[(^\prime\!\nabla_a T_b),\gamma^a]\cr
&\ \ \ +{1\over 4}\;\eta_{ab}\gamma^c[T_c,\gamma^a]\gamma^d[T_d,\gamma^b].
\cr}\eqno(2.14)$$
In the case of $t=0$, i.e. $T=0$, this decomposition obviously
reduces to the
usual Lichnerowicz formula $D^2=\triangle^\nabla + {{\petit 1}\over
{\petit 4}}\; R^\nabla\cdot \eins_S$. Consequently, we call (2.14)
a `generalized Lichnerowicz formula'.
\smallskip
Notice that one has to take into acount that in general
it is impossible to find any $Sk(TM)$-valued one-form
${\hat t}\in \Omega^1(M, Sk(TM))$\ such that the endomorphism part
${\hat T}_X\in \End S$\ of ${\hat T}$\ corresponds to ${\hat t}_X\in
Sk(TM)$\ for all $X\in \Gamma(TM)$. Hence, ${\hat\nabla}^S$\
is generally not induced by any metric connection
${\hat\nabla}$\ on $TM$.
However, if $t\in \Omega^1(M,Sk(TM)$\ is
totally
antisymmetric we obtain the following
\Lemma Let $M$\ be a spin mannifold, $\nabla$\ the Levi-Civita
connection on $TM$\ and $\widetilde{\nabla}$\ defined by
${\widetilde{\nabla}:=\nabla + t}$\ where $t\in \Omega^1(M, Sk(TM))$\
is totally antisymmetric. Then ${{\hat T}=3T}$\ and consequently
${\hat\nabla}^S=\nabla^S+3T$.
\smallskip \rm
This can simply derived from the definiton (2.12) of ${\hat T}$.
\smallskip\rm
\vskip 1.2cm
{\mittel 3. Euclidian Gravity}
\ukneu
\znum=1
\vskip 0.7cm
According to Connes [C] there exists a link between the
usual Dirac operator $D:=\gamma^\mu \nabla^S_\mu$\
on the
spinor bundle $S$\ of a four-dimensional spin manifold $M$\
associated to the Levi-Civita connection
and the euclidian Einstein-Hilbert gravity action via the
Wodzicki residue $Res(D^{-2})$\ of the inverse of $D^2$.
This was explicitly verified in [K2]. Moreover, as already mentioned,
the main theorem of [KW] states that the
the Wodzicki residue
$Res({\hat\triangle}^{-n+1})$\ of any generalized laplacian
${\hat\triangle}$\ acting on sections of an hermitian
vector bundle $E$\ over
an even-dimensional manifold $M$\ with $\dim \;M=2n\ge 4$\
can be identified with
$${{2n-1\over 2}\;\int_M\; * tr\bigl(\Phi_1(x,x,{\hat\triangle})\bigl).
 }$$
Again
$\Phi_1(x,x,{\hat\triangle})$\ denotes the subleading term
of the asymtotic expansion of the heat-kernel
of ${\hat\triangle}$.
In this sense a gravity action can be
defined by
an arbitrary
Dirac operator $\widetilde{D}:=\gamma^\mu\widetilde{\nabla}^S_\mu$\ on $S$\
associated to a metric connection $\widetilde{\nabla}$\
on the base $M$, this means
$${I_{\petit\rm GR}(\widetilde{D}):= -{1\over 2^n}\;
\int_M\;*tr\bigl(\Phi_1(x,x,
\widetilde{D}^2\bigr).}\eqno(3.1)$$
Here $2^n=\dim_\kz S$\ is the complex dimension of the spinor bundle.
By using our generalized Lichnerowicz formula (2.14), we can now
easily
compute ${tr\bigl(\Phi_1(x,x,\widetilde{D}^2)\bigr)}$.
All that remains to be done is
to take traces of $\gamma$-matrices. Thus, we obtain the
\Lemma Let $M$\ be a spin manifold with $\dim\; M=2n$\ even and
$\widetilde{D}\colon \Gamma(S)\rightarrow\Gamma(S)$\ the Dirac operator
on the spinor bundle $S$\ associated to a metric connection
$\widetilde{\nabla}:=\nabla +t$\ as above. Then
$${-{1\over 2^n}\;tr\bigl(\Phi_1(x,x,\widetilde{D}^2)\bigr)=
{1\over 12}\;R^\nabla
+{1\over 2^n}\; \bigl(- t_{abc}t^{abc}+2t_{abc}t^{acb}\bigr)
}\eqno(3.2)$$
with respect to a local orthonormal frame of $TM$.
\smallskip \rm
Note that this result (3.2) holds independently of wether or not
the corresponding Dirac operator $\widetilde{D}$\ is
self-adjoint with respect to the hermitian metric on the
spinor bundle $S$.
In the special
case of the torsion tensor being totally anti-symmetric, (3.2) reduces
to
${-{1\over 2^n}\;tr\bigl(\Phi_1(x,x,\widetilde{D}^2)\bigr)= {1\over
12}\;R^\nabla
 - {3\over 2^n}\;t_{[abc]}t^{[abc]}}$\ as already shown in [KW].
\smallskip
In order to find out wether (3.2) defines a pure (euclidian)
Einstein-Cartan theory\fussnote{${^{(2)}}$}{By an
Einstein-Cartan theory we mean
a gravity theory based on the Einstein-Hilbert action.}
we express the right-hand side of (3.2)
by the scalar curvature $R^{\widetilde{
\nabla}}$\ of $\widetilde{\nabla}$. Using the
well-known formula ${{\bf R}^{\widetilde{\nabla}}=
{\bf R}^\nabla + d^\nabla t + {1\over 2}\;[t\wedge t]}$, where
${{\bf R}^{\widetilde{\nabla}}
\in \Omega^2(M,\End \;TM)}$\ denotes the curvature
of $\widetilde{\nabla}$\ and
$d^\nabla$\ is the exterior covariant derivative corresponding
to the Levi-Civita connection $\nabla$, we can rewrite
(3.2) as follows
$$\eqalign{-{1\over 2^n}\;
tr\bigl(\Phi_1(x,x,\widetilde{D}^2)\bigr) &=
{1\over 12}\;R^{\widetilde{
\nabla}}+ {1\over 12}\; t_{abc}t^{bca} - {1\over 2^n}\;t_{abc}t^{abc}
+{1\over 2^{n-1}}\; t_{abc}t^{acb}\cr
&\ \ \ +{1\over 12}\;{t_{ab}}^b {{t^a}_c}^c - {1\over 12}\;
\nabla_\mu {{t^\mu}_a}^a.\cr}\eqno(3.3)$$
Without additional mater fields, our result (3.2) obviously reduces
to the usual Einstein theory of gravity.
Hence we obtain a result similar to that in [KW].
We also conclude from (3.3) that
it is not possible to obtain a `pure'
Einstein-Cartan theory from the square of
an arbitrary Dirac operator $\widetilde{D}$\ associated to a metric
connection on $M$\ by using the Wodzicki residue.
\vskip 1.2cm
{\mittel 4. Conclusion}
\vskip 0.7cm
In this paper we
proved a generalized version of the well-known
Lichnerowicz formula for
the most general Dirac operator $\widetilde{D}$\
on the spinor bundle of an even-dimensional spin manifold $M$\
associated to a metric connection $\widetilde{\nabla}$\ on
$TM$. Applying this formula,
the subleading term $\Phi_1(x,x,\widetilde{D}^2)$\ of
the heat-kernel expansion of $\widetilde{D}^2$\ is easy to compute.
According to [KW], the trace of this term plays a
key-r${\hat {\rm o}}$le in the definition of a (euclidian)
gravity action
$I_{GR}(\widetilde{D})$\
in the context of the non-commutative differential geometry
introduced by
Connes [C]. This gravity action
can be interpreted as defining a
modified Einstein-Cartan theory.
\smallskip
Finally, we would like to add
that it is also  possible
to derive a combined Einstein-Hilbert/Yang-Mills lagrangian
from an appropriatly defined Dirac operator by using similar
techniques. Moreover, this Dirac operator can be considered as
a deformation of the well-known Dirac-Yukawa operator.
This will be shown in a
forthcoming paper [AT].
\vskip 0.7cm
{\bf Acknowledgements.} We are indebted to T. Sch\"ucker for
his stimulating support. We would also like to
thank Susanne for carefully reading the manuscript. One of us, T.A.
would like to thank E. Binz for his gentle support.
\vskip 0.7cm
\begref
\ref{[AT]} T.Ackermann, J.Tolksdorf,  \sl
 Super-connections and a combined Einstein-Yang-Mills
lagrangian\rm , to appear
\ref{[BGV]} N.Berline, E.Getzler, M.Vergne, \sl Heat kernels and
Dirac operators\rm , Springer (1992)
\ref{[C]} A.Connes, \sl Non-commutative geometry and physics\rm ,
IHES preprint (1993)
\ref{[CL]} A.Connes, J.Lott,\sl Particle models and non-commutaive
geometry,\rm Nucl. Phys. B Proc. Supp. {\bf 18B} (1990) 29-47
\ref{[GHV]} W.Greub, S.Halperin, R.Vanstone, \sl Connections, curvature
and cohomology\rm , vol. 1, Academic press (1976)
\ref{[KW]} W.Kalau, M.Walze, \sl Gravity, non-commutative geometry
and the Wodzicki residue\rm , to appear in Journ. of
Geometry and Physics
\ref{[L]} A.Lichnerowicz, \sl Spineurs harmonique\rm ,
C. R. Acad. Sci. Paris S${\acute {\rm e}}$r. A {\bf 257} (1963)
\ref{[K1]} D.Kastler, \sl A detailed account of Alain Connes' version
of the standard model in non-commutative differential geometry I, II
and III,\rm to appear in Rev. Math. Phys.
\ref{[K2]} D.Kastler, \sl The Dirac operator and gravitation\rm ,
to appear in Com. Math. Phys.

\bye